\documentclass[12pt]{elsarticle}
\usepackage{lineno}

\usepackage{amssymb}
\usepackage{amsmath}
\usepackage{graphicx}
\usepackage{bm}
\usepackage{epsfig}
\usepackage{multirow}
\usepackage{lineno}
\usepackage{slashbox}
\usepackage{array}
\usepackage{colortbl}
\usepackage{lscape}

\journal{Nuclear Instruments and Methods A}
\begin{document}
\begin{frontmatter}
\title{Parametric fitting of  data obtained from detectors with finite resolution and limited acceptance}
\author{N.D. ~Gagunashvili\corref{cor1}}
\ead{nikolai@unak.is}
\cortext[cor1]{Tel.: +3544608505; fax: +3544608998}
\address{University of Akureyri, Borgir, v/Nordursl\'od, IS-600 Akureyri, Iceland }
\begin{abstract}
A goodness-of-fit test for the fitting of a parametric model to
data obtained from a detector with finite resolution and limited
acceptance is proposed. The parameters of the model are found by
minimization of a statistic that is used for comparing
experimental data and simulated reconstructed data.  Numerical
examples are presented to illustrate and validate the fitting
procedure.
\end{abstract}
\end{frontmatter}
\section{Introduction}
The probability density function (PDF) $P(x')$ of a reconstructed
characteristic $x'$ of an event obtained from a detector with
finite resolution and limited acceptance can be represented as
\begin{equation}
P(x')= \frac{\int_{\Omega} p(x)A(x)R(x,x') \,
dx}{\int_{\Omega'}\int_{\Omega} p(x)A(x)R(x,x') \, dx \, dx'},
\label{p1_main}
\end{equation}
where $p(x)$ is the true PDF, $A(x)$ is the acceptance of the
setup, i.e. the probability of recording an event with a
characteristic $x$, and $R(x, x')$ is the experimental resolution,
i.e. the probability of obtaining $x'$ instead of $x$ after the
reconstruction of the event. The integration in (\ref{p1_main}) is
carried out over the domain $\Omega$ of the variable $x$ and the
domain $\Omega'$ of the variable $x'$.

There are two ways of fitting a parametric model $p(x,a_{1},a_{2},
\ldots, a_{l})$ of the true PDF to reconstructed data:
\begin{enumerate}
\item Find an estimate $p_u(x)$ of the true PDF $p(x)$ by solving
an unfolding problem, and then fit a parametric model of the true
PDF $p(x,a_{1},a_{2},$ $ \ldots, a_{l})$  to the unfolded
distribution $p_u(x)$.
    \item Fit a parametric model of the reconstructed PDF, that is,
\begin{equation}
 \frac{\int_{\Omega}p(x, a_{1}, a_{2}, \ldots, a_{l}) A(x)R(x,x') \,
dx}{\int_{\Omega'}\int_{\Omega}p(x, a_{1}, a_{2}, \ldots, a_{l})
A(x)R(x,x') \, dx \, dx'},
\end{equation}
 directly to the reconstructed data.
  \end{enumerate}
These two possibilities have been discussed in \cite{zhigunov,
zechebook}. The acceptance $A(x)$ and the resolution function
$R(x,x')$ must be defined for both methods. In the majority of
cases, they cannot be found analytically, and a Monte Carlo method
is used instead for that purpose. The unfolding (inverse) problem
is known to be an ill-posed problem and cannot be solved without a
priori information about the solution. Any solution of this
problem has, in addition to statistical errors due to the finite
statistics of the experimental data, also systematic errors
related to the use of a priori information. These systematic
errors have an influence on the choice of the parametric model and
the estimation of the parameters in the first method. The second
method is preferable in many cases.

After discretization of the problem, the authors of
\cite{zhigunov} found the acceptance function $A(x)$ and the
resolution function $R(x,x')$, and then used these functions to
fit the parameters of the true distribution. The main disadvantage
of this approach is that the resolution function $R(x,x')$, which
is a matrix after discretization, has rather noisy matrix elements
because in real cases the size of the Monte Carlo sample of events
is of the same order as the size of the experimental sample of
events. Another source of uncertainty is the discretization. Also,
the authors of \cite{zhigunov} did not propose a statistic that
could be used for a goodness-of-fit test.

In \cite{zechebook}, a reweighting procedure for fitting a Monte
Carlo  reconstructed distribution to the reconstructed  data was
proposed. The procedure was presented rather sketchily, and cannot
be repeated even for the  example that was used in
\cite{zechebook} for illustration. There is not a clear
explanation of how the parameters and the errors in them were
calculated. The authors of \cite{zechebook} stated, without proof,
that the statistic used for the fitting of the parameters had a
$\chi^2$ distribution but did not define the number of degrees of
freedom. This makes it impossible to use this statistic for
choosing the best model from a set of alternative parametric
models.

Recently, a test for comparing a weighted histogram
\cite{gagu_comp} that is a generalization of the classical
chi-square test \cite{cramer} has been proposed. In this paper, we
apply the results obtained in \cite{gagu_comp} to develop a
procedure for direct parametric fitting of data obtained from
detectors with finite resolution and limited acceptance.

This paper is organized as follows. In Section 2, a method for
fitting the parameters of the model of the true PDF to the data
and a goodness-of-fit test are proposed. A statistic for comparing
a histogram with unweighted entries and a histogram with weights
that depend on the parameters is used for that purpose. In Section
3, a numerical example that demonstrates how one can estimate the
parameters and the statistical errors in them practically is
presented. A numerical experiment with 10\,000 runs is described
to validate the proposed method.

\section{Parametric fitting of Monte Carlo results to data}

We consider the PDF $P_1(x')$ of a reconstructed characteristic of
experimental events and the PDF $P_2(x')$ of the corresponding
reconstructed characteristic of the Monte Carlo events for the
same detector.

A histogram with $m$ bins for a given PDF $P_1(x')$ is used to
estimate the probability $P_{1i}$ that a random event belongs in
bin $i$:
\begin{equation}
P_{1i}=\int_{S'_i}P_1(x') \, dx', \; i=1,\ldots ,m. \label{p1}
\end{equation}
The integration in (\ref{p1}) is carried out over the bin $S'_i$,
and $\sum_1^m P_{1i}=1$. The histogram can be obtained as the
result of a random experiment with the PDF $P_1(x')$. We denote
the number of random events belonging to the $i$th bin of the
histogram by $n_{1i}$. The total number of events in the histogram
is equal to $n_1=\sum_{i=1}^{m}{n_{1i}}$. The quantity $\hat{P}_i=
n_{1i}/n_1$ is  an estimator of $P_{1i}$ with an expectation value
$\textrm E \,\hat{P_{1i}}=P_{1i}$.

A histogram of the Monte Carlo reconstructed PDF $P_2(x')$ can be
obtained as the result of a random experiment (simulation) that
has three steps \cite{sobol1}:
\begin{enumerate}
\item A random value $x$ is chosen according to a PDF $g(x)$. The
function $g(x)$ is some expected true (initial) distribution
defined in the domain $\Omega$. \item We go back to step 1 again
with probability $1-A(x)$, and to step 3 with probability $A(x)$.
  \item A random value $x'$ is chosen according to the PDF $R(x,x')$.
\end{enumerate}
The events $x'$ are distributed according to the PDF
$P_{2}^{in}(x')$, where
\begin{equation}
P_{2}^{in}(x') = \frac{ \int_{\Omega} g(x)A(x)R(x,x') \, dx }{\int_{\Omega'} \int_{\Omega} g(x)A(x)R(x,x') \, dx \, dx'}.
\label{p2i}
\end{equation}
The quantity $\hat {P_{2i}^{in}}= n_{2i}/n_2$, where $n_{2i}$ is
the number of events belonging to the $i$th bin for a histogram
with total number of events $n_2$, is an estimator of
$P_{2i}^{in}$,
\begin{equation}
P_{2i}^{in} = \int_{S'_i}P_{2}^{in}(x') \, dx', \; i=1,\ldots ,m,
\label{p2i}
\end{equation}
with the expectation value of the estimator equal to
\begin{equation}
\textrm E \,\hat{P_{2i}^{in}}=P_{2i}^{in}.
\end{equation}
It is expected that $A(x)$ and the resolution function $R(x,x')$
for the real setup and for the Monte Carlo simulation will be the
same. This is achieved by adjusting the Monte Carlo simulation
program and by a suitable choice of the domains $\Omega $ and
$\Omega'$ of the variables $x$ and $x'$.

In experimental particle and nuclear physics, step 3 is the most
time-consuming step of the Monte Carlo simulation. This step is
related to the simulation of the process of transport of particles
through a medium and the rather complex registration apparatus.

To use the results of the simulation with an initial PDF $g(x)$ to
calculate a histogram of events distributed according to the PDF
$P_{2}(x')$ with a true PDF $p(x)$, we write the equation for
$P_{2i}$ in the form

\begin{equation}
P_{2i}= \frac{\int_{S'_i} \int_{\Omega} p(x)A(x)R(x,x') \, dx \,
dx'}{\int_{\Omega'} \int_{\Omega} p(x)A(x)R(x,x') \, dx \, dx'} =
\int_{S'_i} \int_{\Omega} w(x)g(x)A(x)R(x,x') \, dx \, dx',
\label{p23}
\end{equation}
where
\begin{equation}
w(x)=p(x)/g(x)\int_{\Omega'} \int_{\Omega} p(x)A(x)R(x,x') \, dx
\, dx' \label{fweight}
\end{equation}
is the weight function. Because of the condition $\sum_iP_{2i}=1$,
we shall call the weights defined above ``normalized weights''
from now on, as opposed to the unnormalized weights
$\check{w}(x)$, which are given by $\check{w}(x)= \textrm{const}
\cdot w(x)$.

The Monte Carlo reconstructed  histogram for the PDF $P_{2}(x')$
can be obtained using reconstructed events for the PDF
$P_{2i}^{in}(x')$ with weights calculated according to
(\ref{fweight}). In this way, we avoid step 3 of the simulation
procedure, which is important in cases where one needs to
calculate Monte Carlo reconstructed histograms for many different
true PDFs.

We denote the sum of the weights of the events in the $i$th bin of
the histogram with normalized weights by
\begin{equation}
W_i= \sum_{k=1}^{n_{2i}}w_i(k), \label{ffweight}
\end{equation}
where $n_{2i}$ is the number of events in bin $i$ and $w_i(k)$ is
the weight of the $k$th event in the $i$th bin.  The quantity
$\hat{P}_{2i}= W_{i}/n_2$ is an estimator of $P_{2i}$ with
expectation value $\textrm E \, \hat{P_{2i}}=P_{2i}$.

A frequently used technique in data analysis is the comparison of
a reconstructed PDF with a Monte Carlo reconstructed PDF through a
comparison of histograms. The hypothesis of homogeneity
\cite{cramer} states that the two histograms represent random
values with identical distributions. This is equivalent to
assuming that there exist $m$ constants $p_1, \ldots, p_m$ such
that $\sum_{i=1}^{m} p_i=1$ and that the probability  of belonging
to the $i$th bin for some  measured value in the experiment and in
the Monte Carlo simulation is equal to $p_i$.

 From here onwards, we use the weighted histogram with \emph{unnormalized} weights and  $\check{W}_i$ denote the sum of the weights of the events in bin $i$. This is convenient because the calculation of normalization factors is
quite problematic in many practical cases.

We introduce the statistics \cite{gagu_comp}
\begin{equation}
 {X}_k^2=\frac{1}{n_1} \sum_{i=1}^m \frac{n_{1i}^2}{p_{i}}-n_1+ \frac{ s_{k}^2}{n_2}+2 s_{k}, \label{sssu}
\end{equation}
where
\begin{equation}
s_{k}=\sqrt{\sum_{i \neq k}{r}_{i} p_{i} \sum_{i \neq k}
{r}_{i}\check{W}_{i}^2/p_{i}} - \sum_{i \neq
 k}{r}_{i}\check{W}_{i} \label{statnorm}
\end{equation}
and
\begin{equation} r_{i}=\sum_{k=1}^{n_{2i}}\check{w}_{i}(k)/\sum_{k=1}^{n_{2i}}\check{w}_{i}^2(k), \label{rat4}
\end{equation}
with the sums extending over all bins $i$ except one bin $k$. In
these equations, the probabilities $p_i$ are unknown, and
estimators of $\hat p_i$ can be found by  minimization of
(\ref{sssu}). We denote by $\hat X_k^2$ the value of ${X}_k^2$
after substitution of the estimators  $\hat p_i$ into
(\ref{sssu}). As shown in \cite{gagu_comp}, the statistic
\begin{equation}
\hat X^2= \textrm {Med}\, \{\hat X_1^2,  \hat X_2^2,  \ldots ,
\hat X_m^2\}\label{stdavu}
\end{equation}
approximately has a $\chi^2_{m-2}$  distribution if the hypothesis of
homogeneity is valid.

We substitute the PDF $p(x)$ by the parametric formula $p(x,a_{1},
a_{2}, \ldots, a_{l})$; the weights of the Monte Carlo events and the statistic $\hat X^2(a_{1},a_{2}, \ldots, a_{l})$ are
then dependent on the parameters. The estimators $\hat a_{1}, \hat
a_{2}, \ldots, \hat a_{l}$ of the parameters $a_{1},a_{2}, \ldots,
a_{l}$ can be found by minimization of this statistic. The
statistic $\hat X^2(\hat a_{1},\hat a_{2}, \ldots, \hat a_{l})$
has a $\chi^2_{m-2-l}$ distribution if the parametric model fits
the data, because $l$ parameters are estimated. It can be used for
a goodness-of-fit test for selection of the best model from a set
of alternative models.

\section{Numerical examples}

In this section two examples are presented to illustrate and validate the fitting procedure.

\subsection{Example 1}

We took the true PDF, as in \cite{zechebook}, to be of the form
\begin{equation}
p(x)=(1+x)/1.5,
\end{equation}
defined on the interval [0,1]. The reconstructed PDF was defined
as
\begin{equation}
P_1(x')= \frac{\int_{\Omega} p(x)A(x)R(x,x') \,
dx}{\int_{\Omega'}\int_{\Omega} p(x)A(x)R(x,x') \, dx \, dx'}, \label{reco}
\end{equation}
with an acceptance function $A(x)=1$ and a resolution function of
the form
\begin{equation}
R(x,x')=\frac{1}{\sqrt{2\mathrm{\pi}}\sigma}\mathrm{exp}
\left(-\frac{(x-x')^2}{2\sigma^2} \right), \label{gauss}
\end{equation}
with $\sigma = 0.3$. The domain of the variable $x$ was taken as
$\Omega = [0,1]$ and that of $x'$ as $\Omega'=[-0.3,1.3]$. A
simulation of events with a PDF $P_1(x')$ was done according to
the algorithm described in the previous section.

For the Monte Carlo reconstructed PDF, we used the initial PDF
$g(x)=1$ and chose the parametrization for the true PDF in the
form
\begin{equation}
p(x,a)\propto 1+ax.
\end{equation}
The Monte Carlo reconstructed PDF was defined as
\begin{equation}
P_2(x')= \frac{\int_{\Omega} p(x,a)A(x)R(x,x') \,
dx}{\int_{\Omega'}\int_{\Omega} p(x,a)A(x)R(x,x') \, dx \, dx'}.\label{mrec}
\end{equation}
Monte Carlo events were  simulated  according to the algorithm
described in the previous section with $g(x)$ as the initial PDF.
Weights of events were calculated according to the formula
$\check{w}(x)=1-ax$. Reconstructed  and Monte Carlo reconstructed samples
were simulated by generating $5 \cdot 10^2$, $5 \cdot 10^3$, and
$5 \cdot 10^4$ events in the first step. We chose 5-bin and 20-bin
histograms and used pairs of histograms with various numbers of
events in the fitting procedure. 10\,000 simulation runs were done
for each case. To investigate the fitting procedure, the following
quantities were calculated:
\begin{itemize}
\item  Average values $\overline{ a}=\sum_1^{1000}\hat a(i)/10
\,000$ of the estimated parameter, where $i$ is the run number.
\item Average statistical errors $\overline{\Delta}$ of the
estimated parameter. For this purpose, the various realizations of
the estimator $ \hat a$ were ordered, and then the positive error
was defined as the minimal interval with lower bound
$\overline{a}$ that contained 34.1345\% of the realizations of $ \hat
a$  and the negative error was defined as the minimal interval
with upper bound $ \overline{a}$ that contained 34.1345\%. \item The
real sizes of the test $\alpha_s$ for a nominal test size
$\alpha=5\%$ were estimated as the fraction of runs that had a
$p$-value lower than 5\%.
\end{itemize}
The program MINUIT \cite{minuit} was used for the minimization of
$\hat X^2(\hat a)$ and for error analysis.

The results of this calculation are presented in Table 1. We may
notice that the estimators are biased in the cases where at least
one histogram is the result of a simulation of $5\cdot10^2$
events. The bias is lower for 20-bin than for 5-bin histograms.
The errors are asymmetric,  and the asymmetry is reduced if the
statistics of the generated events are reduced. The sizes of the
tests $\alpha_s$ are close to the nominal value of $\alpha=5\%$;
see \cite{gagu_comp} for details of the method of comparison.

In the right part of  Table 1, we present results of calculations
for the case where $``\sigma = 0"$, or $R(x,x')=\delta (x-x')$,
which helps us to understand the effect of the resolution
function. The results of calculations for ``infinite" statistics
of the Monte Carlo simulation are also presented. In this case,
the data histograms were fitted by probabilities $p_i(a)$  that
were calculated analytically:
\begin{equation}
p_i(a)=\int_{b_i}^{b_i+1/m}(1+ax) \, dx/\int_0^1(1+ax) \,
dx=\frac{1+ab_{i}+ a/2m}{m+ma/2},
\end{equation}
where $b_i$ is the lower bound of bin $i$. The statistic
\begin{equation}
\sum_{i=1}^m \frac{(n_{1i}-n_1p_i(a))^2}{n_1p_i(a)}
\label{goodness}
\end{equation}
was used to estimate the parameters and for goodness-of-fit tests.
This case represents the best result that can be achieved for
given statistics of the data, and is useful for comparison. The
results presented in Table 1 show that a deterioration of the
resolution leads to an increase in the statistical error of $\hat
a$ and also a bias. We observe an asymmetry in the errors even
in the case of an ``infinite" Monte Carlo simulation. Note that
the statistics (\ref{goodness}) for the estimated values of the
parameters $\hat a$ have a $\chi^2_{m-2}$ distribution if the
experimental histogram is the result of a random experiment with
probabilities $p_i(a), i=1, \ldots, m$ \cite{cramer}.

For the purposes of illustration, we present the results of a
parametric fit of the Monte Carlo results to the data for one of
the cases described above. The numbers of generated  events for
the data and the Monte Carlo simulation were taken equal to $5
\cdot 10^3$, and we used histograms with 20 bins. The result of
fitting with MINUIT was $ \hat a =1.11 ^{+0.30}_{-0.23}$, with
$\hat X^2(\hat a)=11.72$ and the $p$-value equal to 0.82. A
comparison of the histograms of the true PDF with the weighted
histogram of the Monte Carlo true PDF gave $\hat X^2(\hat
a)=18.03$, and the $p$-value was equal to 0.45. Figure 1a shows
the histograms of the reconstructed PDF and of the Monte Carlo
reconstructed PDF, calculated with the weights of the events equal
to $1+1.11x$. Figure 1b shows the histograms of the true PDF and
of the Monte Carlo true PDF.\\
For convenience of visual
comparison, the Monte Carlo histograms have been divided by a
factor $1-\hat a/2=1-1.11/2$ in both figures.

\subsection{Example 2}

We took the true PDF to be of the form
\begin{equation}
p(x) \propto \frac{2}{(x-10)^2+1}+\frac{1}{(x-14)^2+1} \label{testform}
\end{equation}
defined on the interval [4,16]. The reconstructed PDF $P_1(x')$ was defined
according (\ref{reco}) with an acceptance function
 $A(x)=1-(x-10)^2/36$ and a resolution function of
the form defined by formula (\ref{gauss}) with $\sigma = 1.5$. The domain of the variable $x$ was taken as
$\Omega = [4,16]$ and that of $x'$ as $\Omega'=[4,16]$. A
simulation of events with a PDF $P_1(x')$ was done according to
the algorithm described in the previous section.

For the Monte Carlo reconstructed PDF, we used the initial PDF
\begin{equation}
g(x) \propto \check{g}(x)=\frac{3.6}{(x-9)^2+2.25}+\frac{4}{(x-13)^2+4} \label{testform}
\end{equation}
and chose the parametrization for the true PDF in the
form
\begin{equation}
p(x,a_1,a_2,a_3)\propto \check{p}(x,a_1,a_2,a_3)=\frac{a_1}{(x-a_2)^2+1}+\frac{1}{(x-a_3)^2+1}.
\end{equation}
The Monte Carlo reconstructed PDF  $P_2(x')$ was defined according formula (\ref{mrec}).
Monte Carlo events were  simulated  according to the algorithm
described in the previous section with $g(x)$ as the initial PDF.
Weights of events were calculated according to the formula
$\check{w}(x)=\check {g}(x)/\check {p}(x,a_1,a_2,a_3)$. Reconstructed  and Monte Carlo reconstructed samples were simulated by generating  $5 \cdot 10^3$, and $2 \cdot 10^4$ events in the first step. We chose 40-bins histogram.

The result of fitting with MINUIT was $ \hat a_1 =1.87 ^{+0.15}_{-0.13}$, $ \hat a_2 =9.86 ^{+0.05}_{-0.07}$ and $ \hat a_3 =14.05 ^{+0.25}_{-0.20}$ , with
$\hat X^2(\hat a_1,\hat a_2,\hat a_3)=32.00$ and the $p$-value equal to 0.70. A
comparison of the histograms of the true PDF with the weighted
histogram of the Monte Carlo true PDF gave $\hat X^2(\hat
a_1, \hat a_2, \hat a_3)=46.85$, and the $p$-value was equal to 0.15. Figure 2a shows
the histograms of the reconstructed PDF and of the Monte Carlo
reconstructed PDF, calculated with the weights of the events equal
to $\check {g}(x)/\check {p}(x,\hat a_1,\hat a_2,\hat a_3)$. Figure 2b shows the histograms of the true PDF and of the Monte Carlo true PDF.
For convenience of visual
comparison, the Monte Carlo histograms have been divided by a
factor $4\int_4^{16}\check {p}(x,\hat a_1,\hat a_2,\hat a_3)dx $ in both figures.
\newpage
\begin{figure}
\begin{center}$
\begin{array}{cc}
\hspace*{-0.2 cm} \includegraphics[width=3.1in]{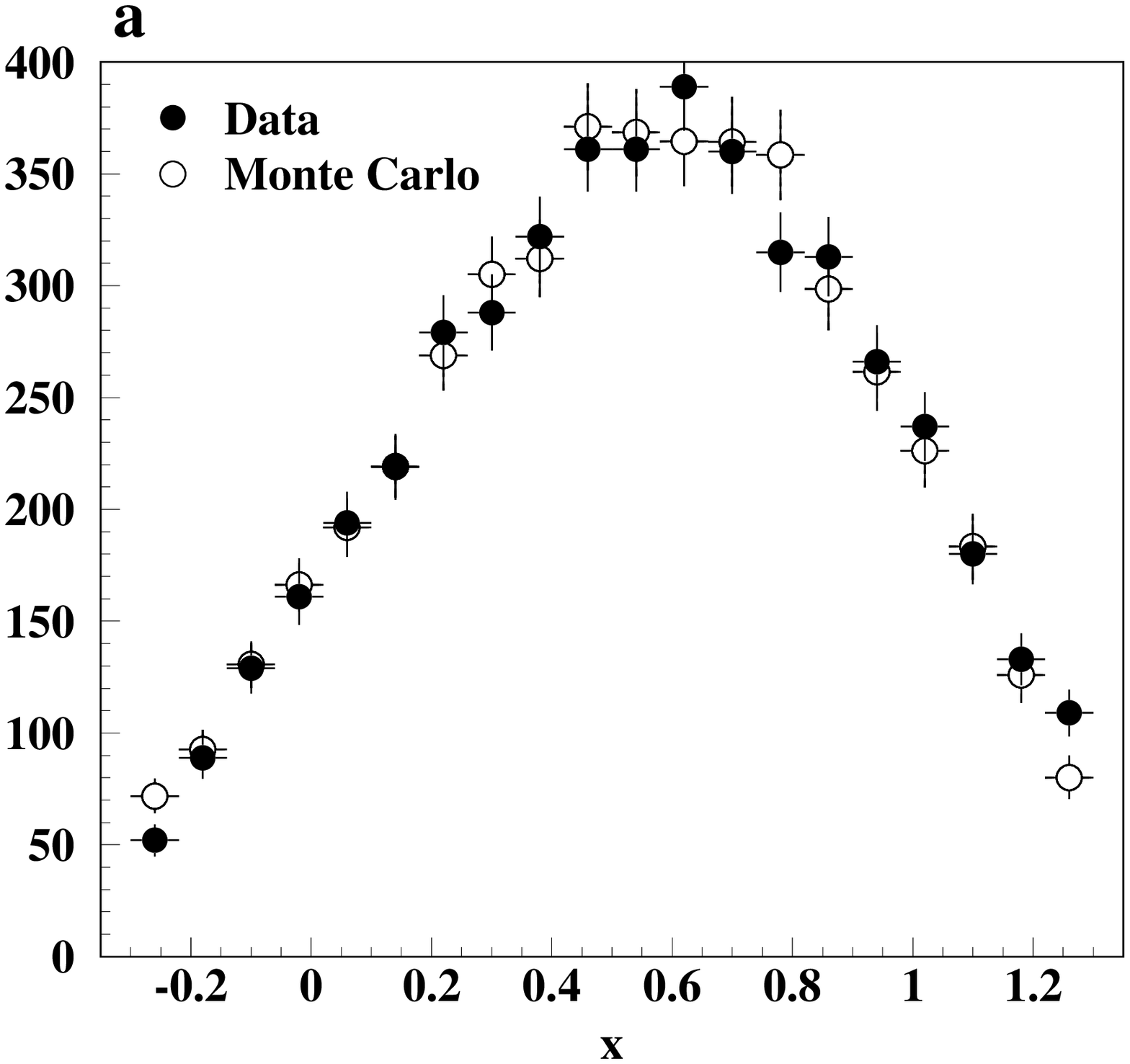}
\hspace*{-1.58 cm}  \includegraphics[width=3.1in]{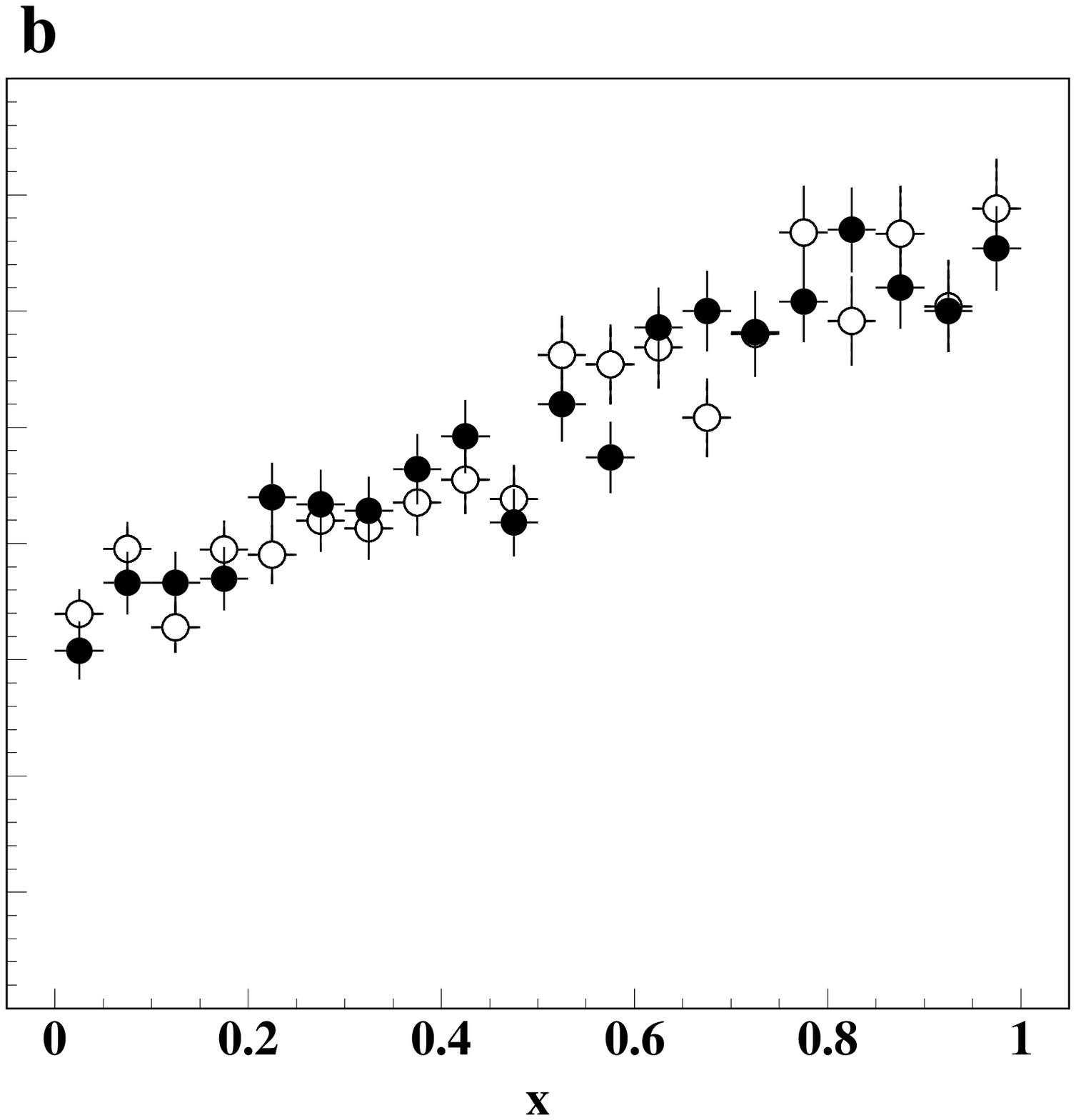}
\end{array}$
\end{center}
\vspace *{-1.2cm}
\parbox{13.5cm}{\caption{{\it Example 1.} Results of parametric fit of
Monte Carlo results to data: (a) histograms of reconstructed PDF
and Monte Carlo reconstructed PDF; (b) histograms of true PDF and
Monte Carlo true PDF.}}
\end{figure}
\begin{figure}
\begin{center}$
\begin{array}{cc}
\hspace*{-0.2 cm} \includegraphics[width=3.1in]{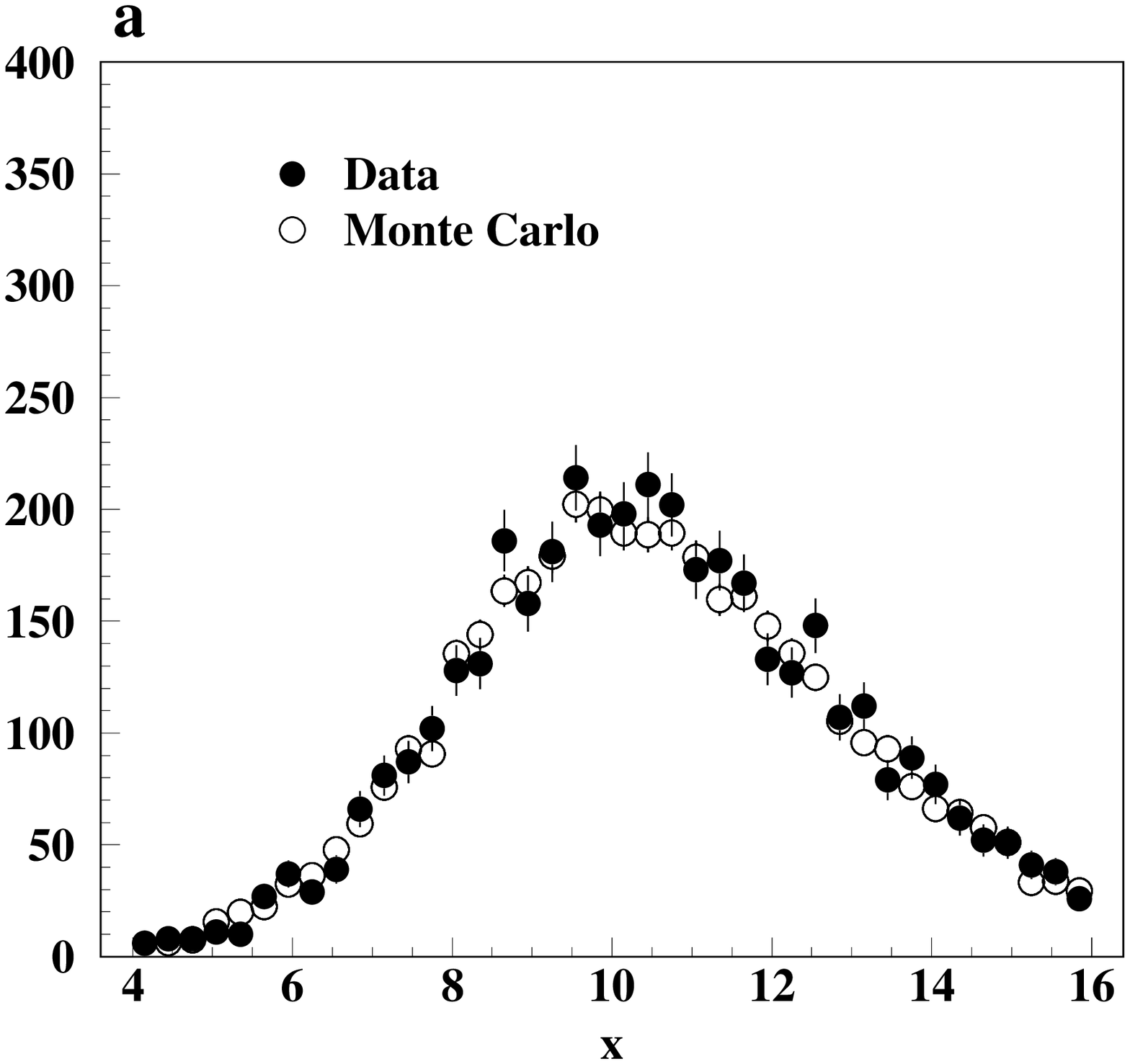}
\hspace*{-1.58 cm}  \includegraphics[width=3.1in]{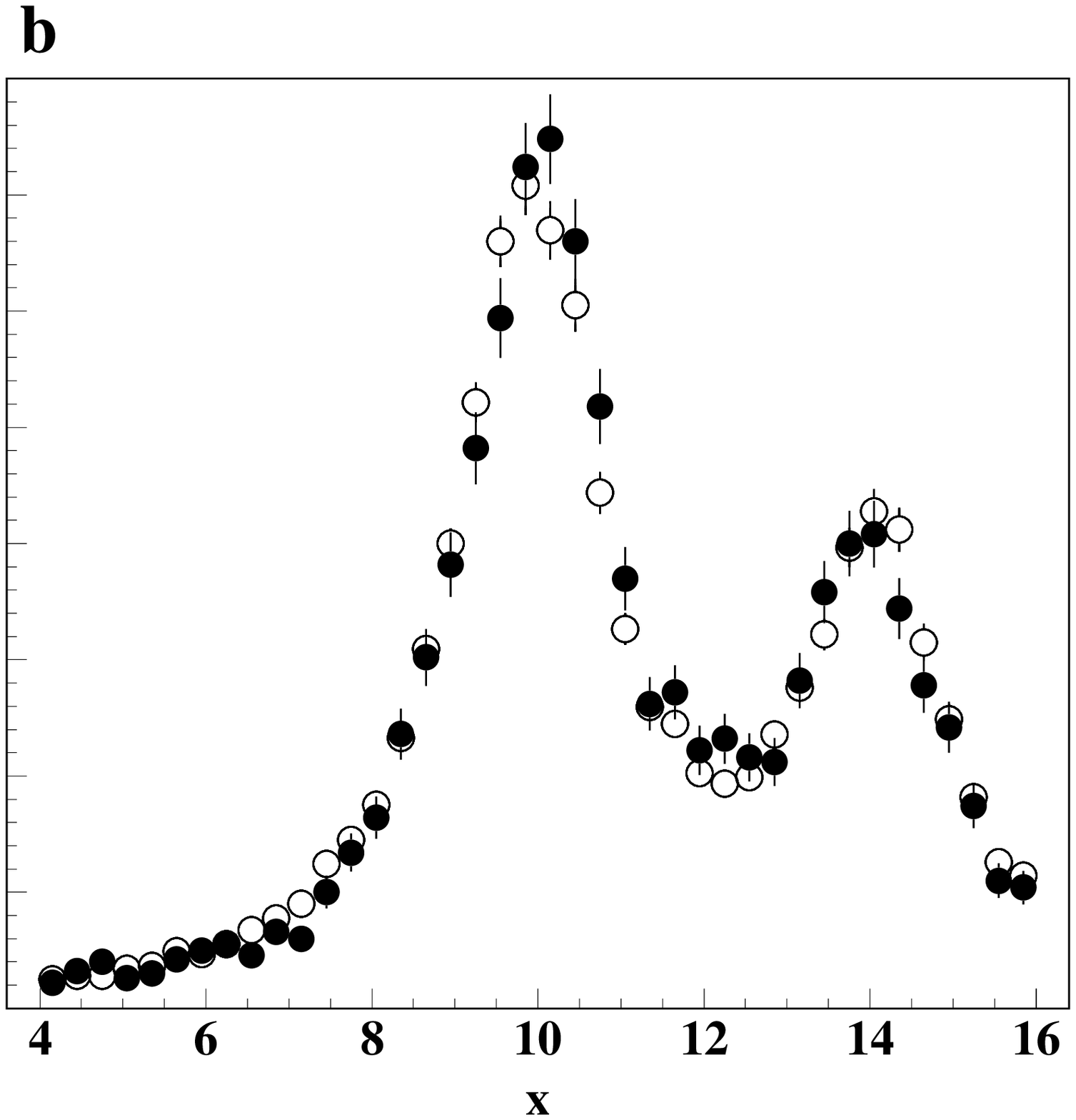}
\end{array}$
\end{center}
\vspace *{-1.2cm}
\parbox{13.5cm}{\caption{{\it Example 2.} Results of parametric fit of
Monte Carlo results to data: (a) histograms of reconstructed PDF
and Monte Carlo reconstructed PDF; (b) histograms of true PDF and
Monte Carlo true PDF.}}
\end{figure}
\newpage
\begin{landscape}
 \begin{table}\footnotesize
\parbox{19.6cm} {\caption{{\it Example 1.} Mean values $\overline a$ of parameter estimates $\hat a$,
mean values $\overline{\Delta}$ of errors of parameter estimates
$\hat a$, and sizes of test $\alpha_s$ for a nominal size of test
$\alpha=5\%$.  Calculations were done for histograms of the
reconstructed PDF and  Monte Carlo reconstructed PDF with various
numbers of generated events $n_{da}$ and $n_{mc}$ for numbers of
bins $m=5$ and $m=20$. The left part of the table presents
calculations for a resolution function with $\sigma=0.3$, and the
right part for $\sigma=0$.}}

\vspace *{0.3 cm}
{\hspace *{0.1 cm} $\sigma=0.3$ \hspace *{8.3 cm}  $\sigma=0$}
\vspace *{0.1 cm}

\begin{minipage}{3.7in}
\begin{center}
\begin{tabular}{|l|l|lll|lll|}
\hline
\multicolumn{2}{|c|}{} & \multicolumn{3}{l|}{$\underline{m=5}  \,\,\,\,\,\qquad n_{mc}$} & \multicolumn{3}{l|}{$\underline{m=20}  \,\,\,\qquad n_{mc}$} \\
\multicolumn {2}{|c|}{$\!\!\!\!\!\!\!\!\!\!\! n_{da}$}  & $5\negthinspace\cdot\negthinspace10^2$ & $5\negthinspace\cdot\negthinspace10^3$ & $5\negthinspace\cdot10^4\negthinspace$ & $5\negthinspace\cdot\negthinspace10^2$ & $5\negthinspace\cdot\negthinspace10^3$ & $5\negthinspace\cdot\negthinspace10^4$ \\
\hline

\multirow{3}{*}{$5\negthinspace\cdot\negthinspace10^2$} & $ {\overline a}$ & 1.29 & 1.16& 1.13& 1.17 & 1.07 & 1.07 \\
 & \multirow{2}{*} {$\overline{\Delta}$}  &  \normalsize$_{+3.13 }$&\normalsize $_{ +1.16}$ &\normalsize $_{ +0.81}$& \normalsize $_{ +1.84}$& \normalsize $_{ +0.85}$& \normalsize $_{ +0.79}$ \\
 & & \normalsize$^{-0.66}$ &\normalsize$^{ -0.52}$&\normalsize$^{ -0.54}$& \normalsize$^{ -0.61}$ & \normalsize$^{ -0.46}$ & \normalsize$^{ -0.45}$\\
 & $\alpha_s$ & 5.2\% & 6.1\% & 6.3\%& 4.8\% & 5.2\%& 5.3\% \\ \hline
\multirow{3}{*}{$5\negthinspace\cdot\negthinspace10^3$} & $ {\overline a}$ & 1.12 & 1.02 & 1.01& 1.11 & 1.02 & 1.00 \\
 & \multirow{2}{*} {$\overline{\Delta}$} & \normalsize $_{+0.93}$&\normalsize$_{ +0.30}$ &\normalsize$_{ +0.23}$ & \normalsize$_{ +0.91}$& \normalsize$_{ +0.28}$ &  \normalsize$_{ +0.20}$\\
 & &\normalsize $^{-0.52 }$ &\normalsize $^{-0.22}$ &\normalsize$^{ -0.17}$ & \normalsize$^{ -0.47} $& \normalsize$^{ -0.22} $& \normalsize$^{ -0.16}$\\
 & $\alpha_s$ & 6.1\% & 5.8\%& 5.5\% & 5.3\% & 5.2\% & 5.8\% \\ \hline
\multirow{3}{*}{$5\negthinspace\cdot\negthinspace10^4$} & $ {\overline a}$ & 1.10 & 1.01& 1.00& l.10 & 1.01 & 1.00 \\
 & \multirow{2}{*} {$\overline{\Delta}$} & \normalsize $_{+0.87}$&\normalsize$_{+0.22}$&\normalsize$_{+0.09}$& \normalsize$_{+0.83}$& \normalsize$_{+0.20}$ & \normalsize$_{+0.08}$\\
 & &\normalsize $^{-0.49}$ & \normalsize $^{-0.17}$ & \normalsize $^{-0.08}$ &\normalsize $^{-0.45}$  & \normalsize $^{-0.16}$ & \normalsize $^{-0.07}$\\
 & $\alpha_s$ &5.4\%  & 6.0\% & 5.7\% & 5.4\% & 5.7\% & 5.6\% \\ \hline
\end{tabular}
\end{center}
\end{minipage}
\vspace{-0.0cm}
\begin{minipage}{4.06in}
\begin{center}
\begin{tabular}{|llll|llll|}
\hline
 \multicolumn{4}{|l|}{$\underline{m=5}  \,\,\,\,\,\qquad \qquad n_{mc}$} & \multicolumn{4}{l|}{$\underline{m=20}  \,\,\,\,\qquad \qquad n_{mc}$} \\
$5\negthinspace\cdot\negthinspace10^2$ & $5\negthinspace\cdot\negthinspace10^3$ & $5\negthinspace\cdot\negthinspace10^4$ & $\infty$& $5\negthinspace\cdot\negthinspace10^2$ & $5\negthinspace\cdot\negthinspace10^3$ & $5\negthinspace\cdot\negthinspace10^4$ &$\infty$ \\
\hline

1.08 & 1.05& 1.04& 1.04 &1.04 & 1.02 & 1.01&1.01 \\
\normalsize $_{+0.68 }$&\normalsize$_{ +0.48}$ &\normalsize$_{ +0.44}$ &\normalsize $_{ +0.43}$&\normalsize $_{ +0.62}$& \normalsize$_{ + 0.44}$& \normalsize$_{ +0.42}$ &\normalsize $_{ +0.40}$\\
\normalsize $^{-0.44}$ &\normalsize$^{ -0.34}$&\normalsize$^{ -0.33}$&\normalsize$^{ -0.29}$ & \normalsize$^{ -0.39}$ & \normalsize$^{ -0.30}$ & \normalsize$^{ -0.28}$&\normalsize$^{ -0.29}$ \\
6.0\% & 6.0\% & 6.4\%& 5.0\% & 5.1\% & 5.6\%& 5.4\% &4.7\%\\ \hline
1.03 & 1.01 & 1.00& 1.00 & 1.02 & 1.01 & 1.00& 1.00 \\
\normalsize $_{+0.40}$&\normalsize$_{ +0.17}$ &\normalsize$_{ +0.13}$ &\normalsize$_{ +0.12}$& \normalsize$_{ +0.38}$& \normalsize$_{ +0.16}$ &  \normalsize$_{ +0.12}$&\normalsize$_{ +0.11}$\\
\normalsize $^{-0.34 }$ &\normalsize $^{-0.15}$ &\normalsize$^{ -0.11}$ &\normalsize$^{ -0.10}$& \normalsize$^{ -0.31} $& \normalsize$^{ -0.14} $& \normalsize$^{ -0.10}$&\normalsize$^{ -0.10}$\\
6.0\% & 5.9\%& 5.8\%& 4.8\% & 5.5\% & 5.6\% & 5.6\% & 4.6\%\\ \hline
1.04 & 1.00& 1.00& 1.00 &l.02 & 1.00 & 1.00 & 1.00 \\
\normalsize $_{+0.40}$&\normalsize$_{+0.12}$&\normalsize$_{+0.05}$&\normalsize$_{ +0.03}$& \normalsize$_{+0.36}$& \normalsize$_{+0.11}$ & \normalsize$_{+0.05}$&\normalsize$_{ +0.03}$\\
\normalsize $^{-0.32}$ & \normalsize $^{-0.11}$ & \normalsize $^{-0.05}$ & \normalsize $^{-0.03}$&\normalsize $^{-0.32}$  & \normalsize $^{-0.11}$ & \normalsize $^{-0.05}$& \normalsize $^{-0.03}$\\
5.7\%  & 5.5\% & 5.8\%& 4.9\% & 5.5\% & 6.0\% & 5.6\%& 5.3\% \\ \hline
\end{tabular}
\end{center}
\end{minipage}
\end{table}
\end{landscape}

\section{Conclusions}

A method of fitting a parametric model to data measured with a
detector with finite resolution and limited acceptance has been
developed. It was developed as an application of a test for
comparing histograms with unweighted entries and histograms with
unnormalized weights proposed in previous work by the present
author. The method demonstrates a new approach to the direct
parametric fitting of experimental data that permits one to
decrease the systematic errors in the estimated parameters. It is
a rather flexible tool for data analysis that can be used with
multidimensional data, and does not have any restrictions on the
configuration of the bins or the domain of the variables
investigated. A goodness-of-fit test has been proposed that can be
used for selection of the best parametric model from a set of
alternative models for describing the data. An evaluation of the
method has been done numerically for histograms with various
numbers of bins and numbers of events. A numerical examples have
been given to demonstrate the use of the method in practice.

\newpage
\end{document}